# Water-Gated Charge Doping of Graphene Induced by Mica Substrates

Jihye Shim,[1,†] Chun Hung Lui,[2,†] Taeg Yeoung Ko,[1] Young-Jun Yu,[2] Philip Kim,[2] Tony F. Heinz[2,3]\*, and Sunmin Ryu[1]\*

[1]Department of Applied Chemistry, Kyung Hee University, Yongin, Gyeonggi 446-701, Korea

[2]Department of Physics, Columbia University, New York, NY 10027, USA

[3]Department of Electrical Engineering, Columbia University, New York, NY 10027, USA

[†]These authors contributed equally to this study

\*E-mail: (T.F.H) tony.heinz@columbia.edu; (S.R.) sunryu@khu.ac.kr

**Abstract**

We report on the existence of water-gated charge doping of graphene deposited on atomically flat mica substrates. Molecular films of water in units of ~0.4 nm-thick bilayers were found to be present in regions of the interface of graphene/mica hetero-stacks prepared by micromechanical exfoliation of kish graphite. The spectral variation of the G and 2D bands, as visualized by Raman mapping, shows that mica substrates induce strong *p*-type doping in graphene, with hole densities of $(9 \pm 2) \times 10^{12}$ cm$^{-2}$. The ultrathin water films, however, effectively block interfacial charge transfer, rendering graphene significantly less hole-doped. Scanning Kelvin probe microscopy independently confirmed a water-gated modulation of the Fermi level by 0.35 eV, in agreement with the optically determined hole density. The manipulation of the electronic properties of graphene demonstrated in this study should serve as a useful tool in realizing future graphene applications.

Charge transfer at interfaces plays an important role in the fundamental properties of physical, chemical and biological systems, as well as in phenomena of practical significance, such as photocatalysis,[1] photosynthesis,[2] photovoltaic processes,[3] light emission by organic light emitting diodes (OLEDs),[4] and charge flow in molecular electronics.[5] In particular, control of the electrical conductivity in organic and low-dimensional semiconductors such as conducting polymers[6] and diamond surfaces[7] has been achieved by charge-transfer doping using molecules with large electron or hole affinity. In the rapidly growing field of graphene research,[8] charge transfer processes are also of prime importance for selective *p*-/*n*-type doping,[9-14] opening electronic band gaps,[15, 16] sensing of individual molecules,[17] and improving the electrical properties of transparent conductive electrodes.[18, 19] It has, for example, recently been shown that chemical treatment of few-layers graphene can produce greatly enhanced sheet conductivity, comparable to that of commercial indium tin oxide (ITO), without degradation of optical transparency.[18]

Despite the technological compatibility and importance of $SiO_2$/Si substrates, graphene supported on these substrates exhibits nanometer-scale ripples[20, 21] and a spatially varying hole density ranging up to $\sim 10^{13}$ carriers/cm$^2$.[22] Several research groups have recently reported that substrate-mediated charge doping can be significantly decreased by modification of the $SiO_2$ surface through silane chemistry.[23-25] However, it is desirable to have separate control of charge transfer process and the surface structure, since the unintentional *p*-type doping has been attributed not only to direct charge transfer with the underlying substrate,[26, 27] but also to the influence of ambient $O_2$ molecules bound to the rippled graphene film.[28-30] From this perspective, alternative insulators like mica[31, 32] and hexagonal boron nitride (h-BN)[33, 34] have attracted great interest because of the decreased roughness, improved charge homogeneity, and enhanced carrier mobility of graphene supported on these substrates. In particular, the morphology of graphene has been found to be free of ripples[35, 36] when deposited on atomically flat substrates of mica[31], a material consisting of alternating aluminosilicate anionic layers and single layers of $K^+$ ions.[37] The graphene/mica interface, which is held together through weak van der Waals forces, provides a unique system for charge transfer since a recent theoretical analysis has predicted that mica may induce either electron or hole doping depending upon the density of surface $K^+$ ions in contact with the graphene.[37] Layer-by-layer formation of water molecular films between graphene and mica also allows control of the interfacial gap with molecular level precision.[32, 38]

In this letter, we demonstrate that mica substrates induce strong and persistent *p*-type doping in graphene, with hole densities of $(9 \pm 2) \times 10^{12}$ cm$^{-2}$. This conclusion is based on analysis by both Raman spectroscopy and scanning Kelvin probe microscopy (SKPM). We also show that ultrathin interfacial water films greatly suppress charge transfer between graphene and the mica substrate. The charge transfer interactions of graphene with the supporting mica substrate identified in this study can be exploited for robust and lasting manipulation of charge carriers in future graphene applications. The method provides a complement to electrical gating[39] and conventional chemical doping as a route for control of carrier densities in graphene.[9-14]

Graphene monolayers were prepared on the surface of freshly cleaved mica by means of mechanical exfoliation. The resulting structures were characterized by Raman spectroscopy, atomic force microscopy (AFM), and SKPM. (See Methods for experimental details.) Figures 1 and 2 present representative topographic maps of such graphene samples obtained by AFM. The images exhibit flat plateaus with areas from 0.001 - 10 μm$^2$. We designate these regions according to their height as $1L_{0WL}$, $1L_{1WL}$, or $1L_{2WL}$. (As discussed below, $nL_{mWL}$ denotes a structure of *n*-layer graphene/*m*-water bilayers/mica.) Raman spectroscopy reveals that both samples in these images are graphene films of single-layer thickness, while the absence of an appreciable D band indicates the high degree of crystalline order of these graphene samples.[40, 41] (See Fig. S1.) As demonstrated by the AFM line profiles in Fig. 1, the height of the plateaus is a multiple of a characteristic thickness of 0.40 ± 0.05 nm, suggesting the presence of ultrathin molecular layers beneath the graphene. According to the recent report of Xu *et al.*,[32] the observed graphene plateaus arise from the formation of water films in discrete units of water bilayers (WL) on the polar mica surface. (See Supporting Information **B** and **C** for a detailed account of the origin of the molecular layers and of the influence of ambient humidity on these layers.) The graphene area near the diagonal edge in Fig. 1a (denoted as $1L_{0WL}$) exhibits a step height of 0.40 nm with respect to the nearby mica surface and is taken to be in direct contact with the substrate. Some samples exhibit extended graphene areas in direct contract with the substrate, as, for example, in Fig. 2a. However, we found that the observed step height across $1L_{0WL}$-mica boundaries varies from 0.3 to 1.5 nm, depending on AFM scanning parameters. This range of apparent step heights is attributed to the chemical and electrostatic differences between graphene and mica (Supporting Information **D**).[32, 42] Considering that the thickness of graphene and of an individual bilayers of water is only ~0.4 nm, the uncertainty in the experimental height data may lead to an incorrect assignment of the structures. We were able to resolve this ambiguity by minimizing the interaction between the AFM tip and samples (Supporting Information **D**), as well as by applying Raman spectroscopic characterization, as discussed below.

We now demonstrate that mica substrates strongly hole dope deposited graphene layers, but that a single interposed bilayer of water effectively suppresses the induced charge doping. This conclusion can be drawn from examination of Raman spectra like those in Fig. 2b. These measurements have been performed along a line crossing the $1L_{0WL}$-$1L_{1WL}$-$1L_{0WL}$ regions of the sample, as indicated in Fig. 2c. The frequencies of the G ($\omega_G$) and 2D ($\omega_{2D}$) Raman bands for $1L_{0WL}$ areas found to be unusually high, 1595 and 2695 cm$^{-1}$, respectively. These values stand in clear contrast to the frequencies of 1580 and 2680 cm$^{-1}$ measured for freestanding graphene[43], as well as to those for graphene on SiO$_2$/Si substrates.[22] For the $1L_{1WL}$ areas, however, both frequencies approach their intrinsic values. Looking more closely at the measured lineshape of the modes, we see that the G band at the $1L_{0WL}$-$1L_{1WL}$ border consists of two Lorentzian components (dashed lines for the fourth spectrum from the top in Fig. 2b) centered at 1583 ($\omega_G^-$) and 1595 cm$^{-1}$ ($\omega_G^+$), respectively. However, the Raman spectra taken within the central $1L_{1WL}$ region in Fig. 2a (*e.g.*, the fifth spectrum from the bottom in Fig. 2b) exhibit predominantly the lower-frequency $\omega_G^-$ component, with only a slight contribution of the high-frequency $\omega_G^+$ contribution (which we attribute to the presence of minor $1L_{0WL}$ regions within the larger $1L_{1WL}$ area). The frequency difference between the G mode Raman frequencies $\omega_G^+$ and $\omega_G^-$ in Fig. 2b implies a hole density of 9x10$^{12}$ cm$^2$ or a shift of the graphene Fermi energy, $\Delta E_F$, of -0.35 eV.[44] We note that while stiffening of the G and 2D modes in graphene can be induced by compressive strain[45, 46], we deduce that charge doping,[44, 47] is the cause of the observed shifts. In addition to the absence of a mechanism to impose compression strain on the graphene, this conclusion is based on the Raman 2D/G intensity ratio and the correlation between the G-mode frequency ($\omega_G$) and linewidth ($\Gamma_G$) discussed below.

The full spatial maps of the Raman response in Figs. 2c-e demonstrate the consistent correlation between charge doping in the graphene layer and the absence of interfacial water layers. Graphene samples in direct contact with mica show the aforementioned blueshifts for both the G and 2D modes, while graphene regions supported on islands of single bilayers of water all exhibit Raman shifts $\omega_G$ and $\omega_{2D}$ near their intrinsic values. (Also see the blue squares in Fig. 3a for the statistical correlation between the values of $\omega_G$ and $\omega_{2D}$ at differing points in the spatial image.) The ratio ($I_{2D}/I_G$) of the 2D-to-G integrated intensity is also known to be sensitive to charge doping. The strength of Raman 2D mode decreases with increasing charge density;[48] while the strength G mode remains rather constant except until the shift of the Fermi energy |$\Delta E_F$| begins to approach half the photon energy of Raman excitation laser,[49] a regime that is not relevant for the current studies. As shown in Fig. 2e, the 1L$_{0WL}$ regions exhibit smaller values for $I_{2D}/I_G$ than are found in the 1L$_{1WL}$ regions of the sample. We note that the G band spectra from the entire graphene area can be decomposed into two peaks with fixed values of $\omega_G^+$ and $\omega_G^-$ (blue squares in Fig. 3b), but varying strengths. This observation indicates that the local charge density assumes one of two discrete values (~0 and 9x10$^{12}$ cm$^2$), rather than varying continuously throughout the entire graphene area.

We have also confirmed that the water-layer control of hole doping is a general phenomenon. It was found to occur for graphene samples of differing layer thickness resting directly on the mica substrate or on interfacial water layers of differing thickness. We summarize Raman mapping data on several samples by plotting correlations for differing parameters in the Raman spectra in Fig. 3. (See Supporting Information **E** for AFM images.) Fig. 3a displays the correlation between values of $\omega_G$ and $\omega_{2D}$. A positive correlation between these quantities is observed, with an average slope ($\Delta\omega_{2D}/\Delta\omega_G$) for the deviations of 0.98 ± 0.01 (dashed line in Fig. 3a). This result agrees well with that obtained for electrically hole-doped graphene (green squares and line in Fig. 3a).[47] The slope, however, is much less than the value of 2.8 - 3.0 that would be expected for graphene under biaxial tensile or compressive strain.[46, 50] The inverse correlation between $\omega_G$ and $\Gamma_G$ displayed in Fig. 3b, which arises from elimination of nonadiabatic phonon decay channels for |$\Delta E_F$| > $\hbar\omega_G/2$, is also consistent with a theoretical prediction for charge-doped graphene (dashed line in Fig. 3b).[44, 47] We note that the data for the three samples (**M04**, **M05** and **M11**) in Fig. 3b display a division into two groups of behavior for the $\omega_G$-$\Gamma_G$ parameters. This result is consistent with the topographic features of the samples, since they contain both $1L_{0WL}$ and $1L_{mWL; m>0}$ graphene areas (Fig. 2a & S4). The other two samples (**M08** and **M09**), in contrast, show only one localized group of parameters, which correlates directly with the observed topography of these samples (Fig. 1 & S4). The graphene layer of **M09** (~100 μm$^2$), for example, consists of ~60% of $1L_{2WL}$ and ~40% of $1L_{1WL}$, with negligible $1L_{0WL}$ regions (Fig. 1). Since water layers decouple the whole graphene area from the substrate, both G and 2D modes have nearly their intrinsic values across the entire graphene sheet. Fig. 3c shows the consistency of both the peak intensity ratios, $I_{2D}/I_G$, and of the peak height ratios, $H_{2D}/H_G$, for the different samples.

Water-gating of the mica-induced doping was also observed directly in surface potential maps obtained by SKPM (Fig. 4a).[51] Since the spatial variation in the surface potential ($\Delta V$) of an electrically connected graphene flake can be related to its work function variation ($\Delta \Phi$) by $\Delta \Phi = -e\Delta V$,[52] the map provides a direct measure of local variation in the work function ($\Phi$) or the Fermi level ($E_F$).[53] While the $1L_{1WL}$ graphene and the nearby 5-layer (5L) graphene exhibit similar measured potentials, the potential values for the $1L_{0WL}$ graphene areas are a few hundred mV lower (Fig. 4b). The surface potential distribution of the single-layer graphene area consists of two Gaussian components, with an average difference of 0.35 V (Fig. 4c). The 5L graphene area is 0.10 V higher than the $1L_{1WL}$ graphene region. Assuming that the 5L graphene has the same work function of 4.6 eV as bulk graphite,[54, 55] then the work functions of the $1L_{0WL}$ and $1L_{1WL}$ regions are 5.05 and 4.70 eV, respectively. This again confirms that mica strongly hole dopes graphene and that one bilayer of water virtually blocks the doping. The difference in the work function between the $1L_{0WL}$ and $1L_{1WL}$ graphene regions is 0.35 eV, which is in good agreement with the value determined by Raman spectroscopy.

With respect to stability of the doping effects described above, we note that the mica-induced charge doping remained unchanged when the sample was held under ambient conditions for a time period of one year. From the point of view of applications, this long-term stability, corresponding to permanent modification of the charge density in graphene, offers advantages compared with charge doping induced by electrical gating, which requires a continuously applied potential, or by intercalants, which are typically susceptible to further reaction or desorption.[56]

The observed charge transfer behavior also sheds light on the electronic structure of the graphene-mica interface. The direction of charge migration at a solid-solid interface is largely determined by the work functions and electron affinities of the materials,[4, 52] although interfacial chemical interactions may also play a role.[57] Hole-doping indicates that the electron affinity of mica surface is larger than the work function of graphene. The work function of uncharged graphene, which is equivalent to its electron affinity (because of the lack of a band gap), has been experimentally determined to be 4.57 ± 0.05 eV.[53] However, a recent theoretical study predicts that the counterparts for the mica surface could vary over a range as wide as 2.8 - 9.1 eV. The values would depend on the surface density of K atoms, with a deficiency leading to a higher electron affinity and work function.[37]

While both faces formed by cleaving mica crystals are thought to have equal densities of surface K atoms ($n_0/2$, where $n_0$ is K atom density for single K atom layers in bulk mica) on a large length scale,[58] the mechanical perturbation imposed by cleaving and the weak binding of K atoms are likely to cause local inhomogeneity in the distribution of K atoms at the surface, leading to random (and as yet poorly understood) surface domains with varying K atom density.[58, 59] The slight variation of $\omega_{G+}$ and $\omega_{G-}$ among the studied samples (Fig. 3b) may be explained by such variability in the K atom density. In particular, the substrate of sample **M11** appears slightly more electronegative than the others in Fig. 3b, so that even the graphene area on the water layers has a non-negligible hole density, as judged from the blueshift of $\omega_{G-}$. Nevertheless, the doping level of graphene directly supported on mica shows a narrow distribution. Furthermore, the absence of electron-doped graphene/mica samples is not consistent with an inhomogeneous distribution of K atoms. These observations suggest that mica substrates cleaved under ambient conditions (at varying levels of relative humidity) exhibit work functions within a fairly narrow range. According to the aforementioned theoretical study,[37] such homogeneity in the work function among different samples implies a homogeneous distribution of surface K atoms with a density of $n_0/2$, which may be achieved during cleavage because of thermodynamic stability.[60] Since such "electroneutral" mica surfaces are predicted to electron-dope graphene, unlike the strong hole-doping observed in this work, the mica surfaces are presumed to undergo surface relaxation or modification immediately after cleavage. It has long been known that the susceptibility of freshly cleaved mica surface to surface charging is highly dependent on the gas environment.[61] Further, the crystallization of certain salts on mica surface is affected by the gas atmosphere where cleavage is carried out.[61] Some of the environmental sensitivity may be related to the high mobility[62] of surface K ions and the adsorption of gas molecules in the presence of water. The precise mechanism by which water bilayers, when present between the deposited graphene and the mica surface, suppress doping of the graphene films is unclear. The effect may arise from the existence of tunneling barrier from the water bilayer. In addition, the degree of the charge transfer would be affected by the dipole moment of the water bilayers,[63] the direction and magnitude of which are not independently known for our experimental conditions.

In conclusion, we have demonstrated that graphene in direct contact with freshly cleaved mica substrates is strongly hole-doped, with a carrier density of $(9 \pm 2) \times 10^{12}$ cm$^{-2}$ through permanent charge transfer to the substrate. The narrow distribution of the hole density suggests that mica undergoes surface relaxations or modifications immediately after cleaving. Furthermore, bilayer films of interfacial water of 0.4 nm thickness were found to suppress the charge transfer to a high degree.

These results provide a route to precise molecular control of the charge density in graphene. The findings, it is hoped, will contribute to our ability to manipulate the electronic properties of graphene for diverse applications.

**Methods**

*Sample preparation*

Our graphene samples were prepared by mechanical exfoliation of kish graphite (Covalent Materials, Inc.) onto mica substrates (Ted Pella, Grade V1 muscovite mica).[31] Fresh mica surfaces were also prepared by exfoliation. To test the effect of ambient water vapor on formation of interfacial water layers,[32] we varied the ambient relative humidity (RH) and temperature, but found no meaningful correlation between these conditions and the presence of interfacial water layers. (See Supporting Information C.) Following an initial screening of the prepared samples by optical microscopy, the numbers of layers, structural quality, and charge density of deposited graphene sheets were characterized by Raman spectroscopy.[22, 41]

*Raman spectroscopy*

The micro Raman setup consists of an optical microscope (Olympus, IX-71), a spectrograph (Princeton Instruments, SpectraPro 2300i, focal length of 300 mm), and a liquid-nitrogen cooled CCD detector (Princeton Instruments, SPEC-10). All the Raman spectra were obtained in a back scattering geometry using a 40x objective lens (NA = 0.60) under ambient conditions. An Ar-ion laser operating at a wavelength of 514.5 nm was used as the excitation source for the Raman measurements. The spectral resolution, as determined by width of the Rayleigh scattering line, was 3.0 cm$^{-1}$. For the two-dimensional Raman spatial maps, spectra were obtained every 0.5 μm or 1 μm using an *x-y* motorized stage (Mad City Labs, MicroStage). Under typical measurement conditions, we used a laser power of 1 - 3 mW focused onto to spot of 0.5 μm diameter. There was no evidence of laser-induced damage or modification of the sample during the measurements.

*AFM and SKPM*

The topography of the graphene samples on mica was characterized by atomic force microscopy (AFM). We obtained contact and non-contact mode images on a Park Systems XE-70 AFM and tapping mode images on a Digital Instrument Dimension 3100 AFM. All AFM images were collected under ambient conditions. The AFM scanning parameters were found to affect the apparent height differences at graphene-mica boundaries. (See Supporting Information D.) For scanning Kelvin probe microscopy (SKPM) measurements, an AC voltage with an amplitude of $V_{AC}$ = 0.5 V at a frequency of $f$=17 kHz was applied to a Cr/Au coated probe (MikroMasch, NSC14/Cr-Au). The contact potential difference (CPD, $V_{CPD}$) of samples with respect to the tip was determined by applying a DC feedback voltage ($V_{DC}$) to cancel the component of the electrostatic force between the sample and tip oscillating at frequency $f$; $F_f \propto (\partial C/\partial z)(V_{CPD} - V_{DC})V_{AC}$, where ($\partial C/\partial z$) is the gradient of the tip-sample capacitance along the surface normal. To avoid topographic artifacts, the SKPM measurements were carried out during constant-height reverse scans, with the tip lifted 20 nm above the surface plane that was mapped during each of the forward scans in the non-contact AFM topography image.

**Supporting Information Available**

Raman spectra of 1L and 2L graphene supported on mica, note on the molecular origin of the interfacial layers, effects of relative humidity on formation of interfacial water layers, effects of scanning parameters on AFM topographic data, non-contact AFM height images of the samples employed in Fig. 3. This material is available free of charge via the Internet at http://pubs.acs.org.

**Acknowledgment**

This research was supported by Basic Science Research Program through the National Research Foundation of Korea (NRF) funded by the Ministry of Education, Science and Technology (2011-0003374 & 2011-0010863) (S.R.). Research at Columbia University (T.F.H. and P.K.) was carried out as part of the Energy Frontier Research Center funded by the U.S. Department of Energy, Office of Basic Energy Sciences under award number DE-SC0001085. We thank Hyuksang Kwon, Dong Seok Hwang, Sang Myung Lee, and Seong Ho Kim for productive discussion and their efforts in preparing samples.

**Figures**

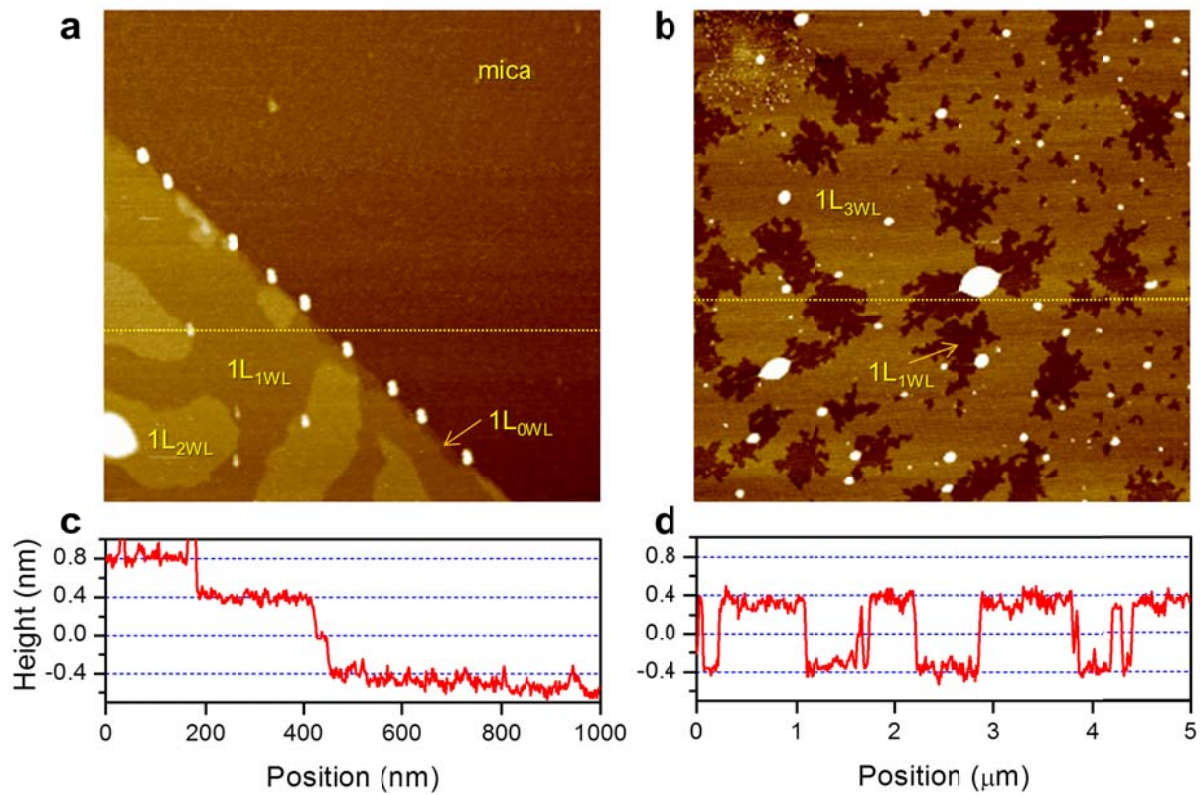

**Figure 1. Interfacial water layers trapped between graphene and mica. a&b**, AFM height images of two representative samples of graphene supported on mica substrates (left: **M09**, right: **M08**) obtained in non-contact mode. **c&d**, Height line profiles taken along the yellow dotted lines in **a** and **b**, respectively. The label $1L_{0WL}$ indicates an area of single-layer graphene in direct contact with mica, while $1L_{1WL}$, $1L_{2WL}$ and $1L_{3WL}$ represent areas of graphene decoupled from mica by 1 to 3 interfacial water bilayers (WL) units, respectively.

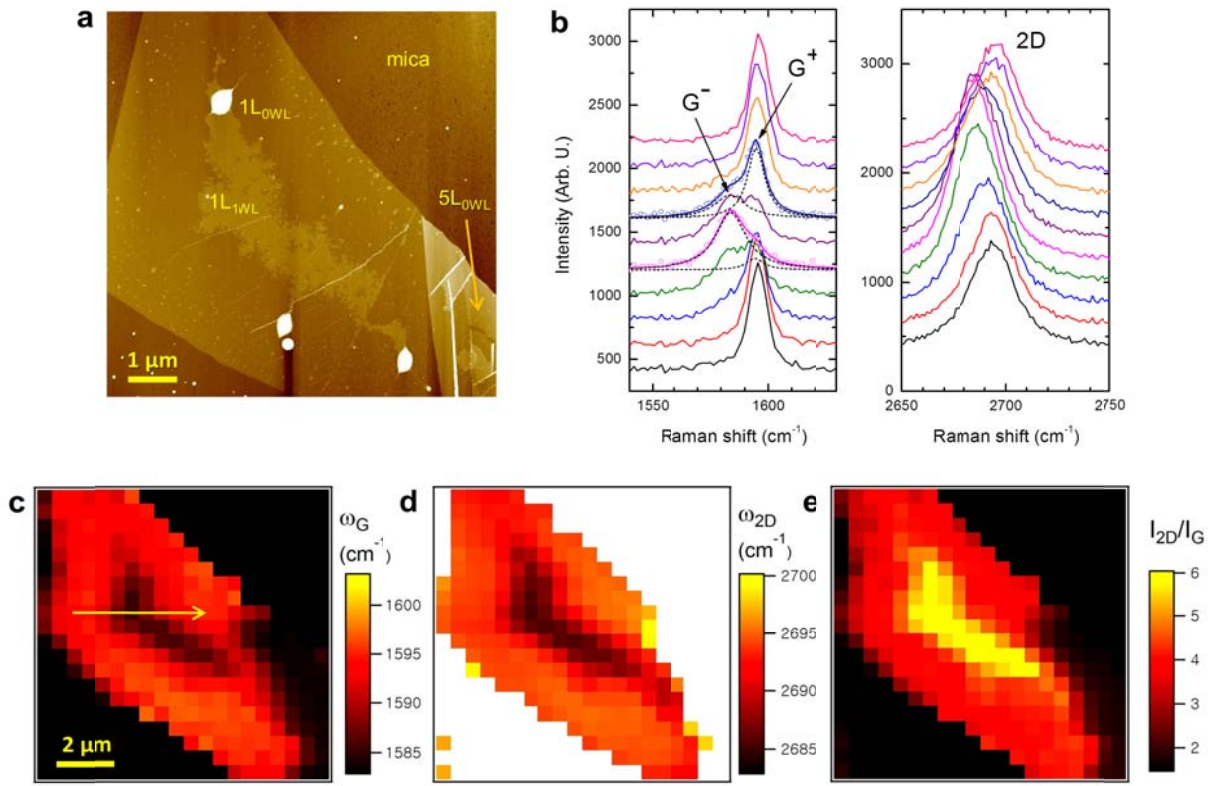

**Figure 2. Raman spectroscopic characterization of water-gated hole doping in graphene on mica. a**, AFM height image of graphene on mica (**M04**) obtained in the non-contact mode. **b**, A series of G and 2D band Raman spectra (solid lines) taken for graphene on mica along the arrow in **c** (from bottom to top). The G band of the fifth and seventh spectra in **b** was fitted with two Lorentzian components, G$^-$ and G$^+$ (dashed lines): the solid line through the data (circles) is the sum of the two. **c**-**e**, Spatial Raman maps of G and 2D bands taken for graphene on mica substrate: the G-band frequency, $\omega_G$ (**c**); the 2D-band frequency, $\omega_{2D}$ (**d**); the integrated intensity ratio of 2D to G band, $I_{2D}/I_G$ (**e**). As a representative value for the G band frequency in **c**, the first moment ($\omega_G^{1st}$) of the G band spectra was used instead of the peak frequency to account for the asymmetry arising from the presence of two components (G$^-$ and G$^+$). The multilayer areas are not clearly resolved in the Raman maps since their values are out of the specified ranges.

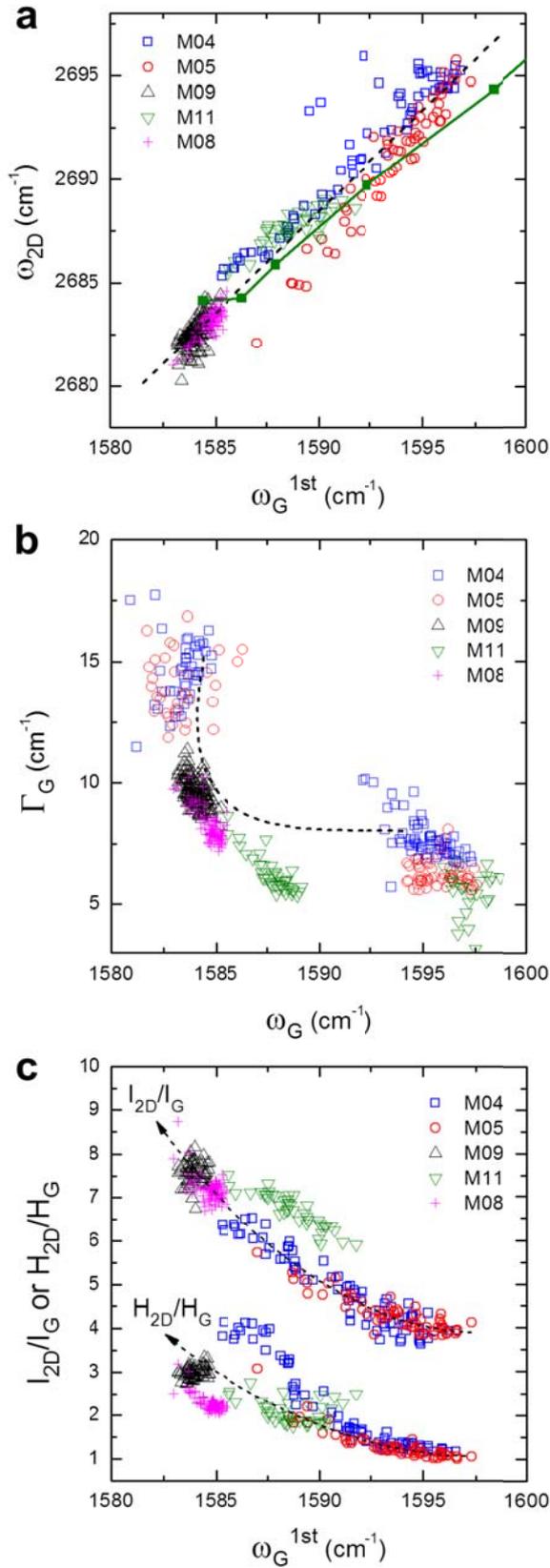

**Figure 3. Statistical representation of spectroscopic features obtained from Raman mapping of various samples**: **a**, Correlation between $\omega_G$ and $\omega_{2D}$. A linear fit (dashed line) to the entire data set yields a slope ($\Delta\omega_{2D}/\Delta\omega_G$) of 0.98 ± 0.01. The green square and solid line represents electrically hole-doped graphene (Ref. 47). **b**, Correlation between $\omega_G$ and line width ($\Gamma_G$) of the G band. The dashed line represents a theoretical prediction for charge-doped graphene (Ref. 22). **c**, Correlation between $\omega_G$ and the 2D/G integrated intensity ($I_{2D}/I_G$) and height ($H_{2D}/H_G$) ratios, where the dashed lines serve as guides to the eye. As representative G band frequencies in **a** & **c**, the first moment, $\omega_G^{1st}$, was used to account for the asymmetric spectra from the presence of the two components (G$^-$ and G$^+$). Each data point corresponds to one pixel of Raman maps, as in Fig. 2 for **M04**. The noticeable deviation of **M04** in $H_{2D}/H_G$ from the other samples in **c** is due to the fact that the splitting into G$^-$ and G$^+$ bands reduces the overall peak height, yielding unusually high $H_{2D}/H_G$ ratios.

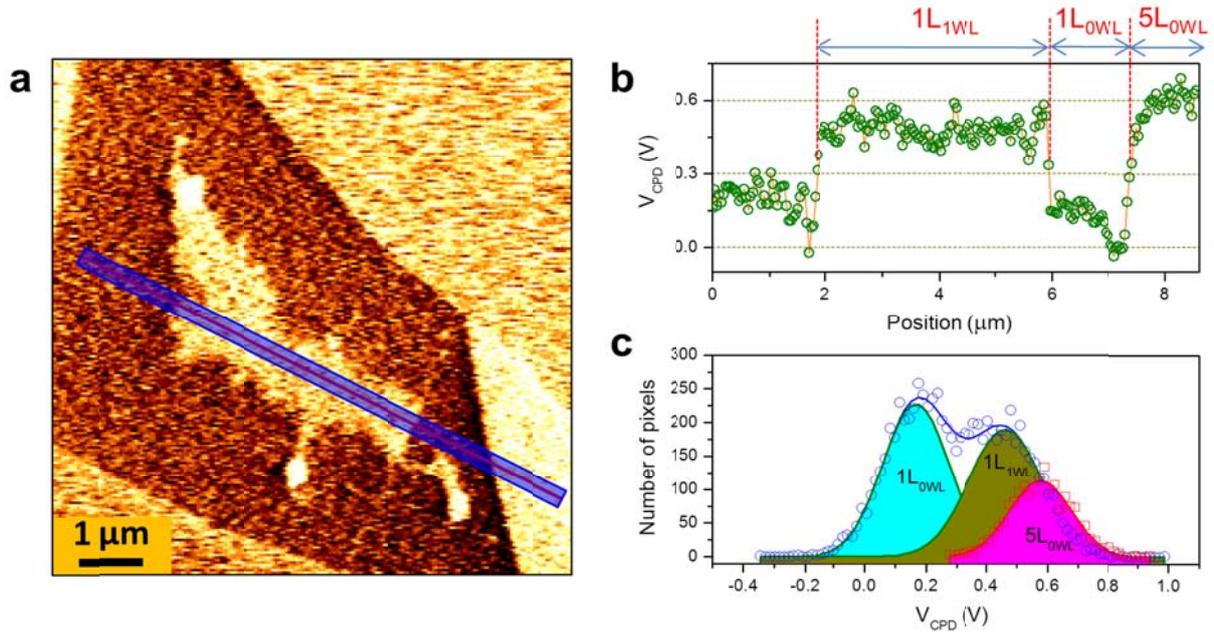

**Figure 4. Electrostatic mapping of water-gated doping in graphene on mica. a**, Spatial map of surface potential obtained for **M04** (see Fig. 2a for its height image) by scanning Kelvin probe microscopy (SKPM). The potential is given as contact potential difference ($V_{CPD}$) with respect to Au/Cr-coated ATM tips. **b**, Line profile of $V_{CPD}$ measured along the red line and averaged in the blue-shaded box in **a**. **c**, Distributions of $V_{CPD}$ taken for 1L (blue circles) and 5L (red squares) graphene areas. The 1L data are fit by two Gaussian functions (green lines) corresponding to $1L_{0WL}$ and $1L_{1WL}$ areas.

**TOC Figure**

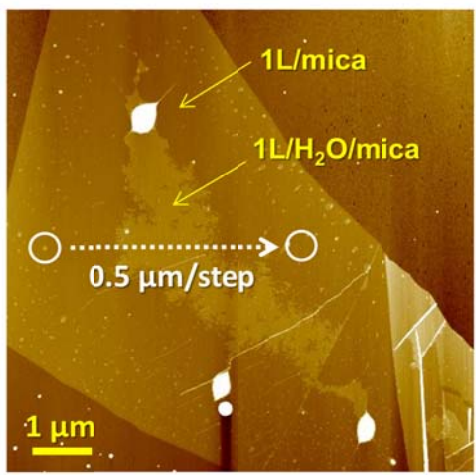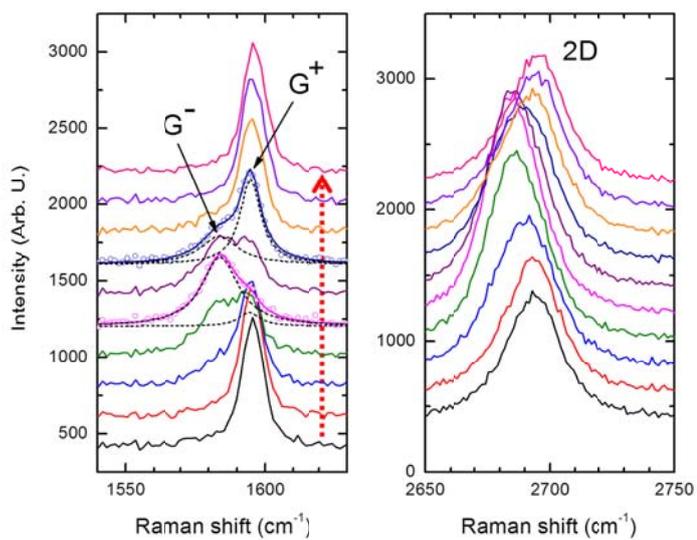